\documentclass[
    ,final            % use final for the camera ready runs
  ]
  {aipproc}

\layoutstyle{6x9}

\newcommand{\nn}{\nonumber}
\newcommand{\be}{\begin{equation}}
\newcommand{\ee}{\end{equation}}
\newcommand{\ba}{\begin{eqnarray}}
\newcommand{\ea}{\end{eqnarray}}

\def\gev{~{\rm GeV}}

\def\als{\alpha_{\rm s}}

\def\qbq{q\overline{q}}

\newcommand{\ov}[1]{\overline#1} 

\newcommand{\req}[1]{(\ref{#1})}
\newcommand{\da}{{distribution amplitude}}

\begin{document}

\title{The gluon contents of the $\eta$ and $\eta^\prime$ mesons }

\author{P.\ Kroll}{
  address={Fachbereich Physik, Universit\"at Wuppertal,\\ 
D-42097 Wuppertal, Germany\\
Email: kroll@physik.uni-wuppertal.de}
}

\begin{abstract}
It is reported on a leading-twist analysis of the $\eta - \gamma$ and
$\eta^\prime - \gamma$ transition form factors. The analysis allows
for an estimate of the lowest Gegenbauer coefficients of the quark
and gluon \da{}s.
\end{abstract}

\maketitle

%%%%%%%%%%%%%%%%%%%%%%%%%%%%%%%%%%%%%%%%%%%%
%% MAINMATTER
%%%%%%%%%%%%%%%%%%%%%%%%%%%%%%%%%%%%%%%%%%%%

One of the simplest exclusive observables is the form factor
$F_{P\gamma^{(*)}}$ for the transitions from a real or virtual photon
to a pseudoscalar meson $P$. Its behaviour at large momentum transfer
is determined by the expansion of a product of two electromagnetic
currents about light-like distances. The form factor then factorizes 
\cite{bro80} into a hard scattering amplitude and a soft matrix element,
parameterized by a process-independent meson distribution amplitude{} 
$\Phi_P$. For space-like momentum transfer the form factor can be 
accessed in $e^+ e^-\to e^+ e^-P$. Such measurements have been carried 
through for quasi-real photons by CLEO \cite{CLEO} and L3 \cite{L3}. 
From the data on the form factors one may extract information about 
the meson distribution amplitudes by fitting the theoretical results
to the experimental data. Here, in this talk, it is reported on recent 
attempts \cite{DKV1,kornelija} to perform such analyses to
leading-twist NLO accuracy in the cases of the $\eta$ and
$\eta^\prime$ mesons. 

As the valence Fock components of the $\eta$ and $\eta^\prime$ mesons
$SU(3)_F$ singlet and octet combinations of quark-antiquark parton
states are chosen
\be
|\qbq_1\rangle = |u\ov{u} + d\ov{d} + s\ov{s}\rangle/\sqrt{3}\,, \qquad
|\qbq_8\rangle = |u\ov{u} + d\ov{d} - 2s\ov{s}\rangle/\sqrt{6}\,.
\ee
In addition the two-gluon Fock state, $|gg\rangle$, is to be
taken into account which also possesses flavour-singlet quantum
numbers and contributes to leading-twist order. 
Associated to each valence Fock component of the meson $P$ is a \da{} 
denoted by $\Phi_{Pi}$ ($i=1,8$) and $\Phi_{Pg}$. The \da s possess 
Gegenbauer expansions \cite{bro80} 
\ba
\Phi_{Pi}(\xi,\mu_F)&=& \frac32(1-\xi^2)\, \Big[1+ \sum_{n=2,4,\cdots}
  B_{Pn}^{(i)}(\mu_F)\; C_n^{3/2}(\xi)\Big] \,, \nn\\
\Phi_{Pg}(\xi,\mu_F)&=& \frac{1}{16}(1-\xi^2)^2\, \sum_{n=2,4,\cdots}
  B_{Pn}^{(g)}(\mu_F)\; C_{n-1}^{5/2}(\xi) \,, 
\label{eq:gegenbauer}
\ea
where $\xi=2x-1$, and $x$ is the usual momentum fraction carried by the
quark inside the meson. The Gegenbauer coefficients, $B_{Pn}$, which 
encode the soft physics, evolve with the factorization scale $\mu_F$ 
according to the relevant anomalous dimensions. The essential point is 
that the singlet and gluon coefficients mix under evolution
\be
B_{Pn}^{(1)}(\mu_F) \leftrightarrow B_{Pn}^{(g)}(\mu_F) \,,
\ee
and that all coefficients evolve to zero for asymptotically large
factorization scales. Hence
\be
\Phi_{Pi}\to \Phi_{AS}=\frac32(1-\xi^2)\,, \qquad \Phi_{Pg} \to 0\,,
\quad {\rm for}\, \mu_F\to \infty\,.
\ee
It is important to note that the gluon \da{} goes along with the following
projector of a state of two incoming collinear gluons (colours $a$,
$b$, Lorentz indices $\mu$, $\nu$ and momentum fractions $x$, $1-x$)
onto a pseudoscalar meson state
\be
{\cal P}_{\mu\nu ,ab}^g = \frac{i}{2}\, \sqrt{\frac{C_F}{n_f}}\,
\frac{\delta_{ab}}{\sqrt{N_c^2-1}}\,\frac{\epsilon_{\perp
      \mu\nu}}{x(1-x)}\,.
\ee  
The anomalous dimensions have to be normalized accordingly \cite{kornelija}. The
components of the transverse polarization tensor are $\epsilon_{\perp
12}=-\epsilon_{\perp 21}=1$ and zero for all others.

\begin{figure}
\parbox{\textwidth}{
\begin{center}
\includegraphics[width=3.6cm, bb=200 470 440 660,clip=true]{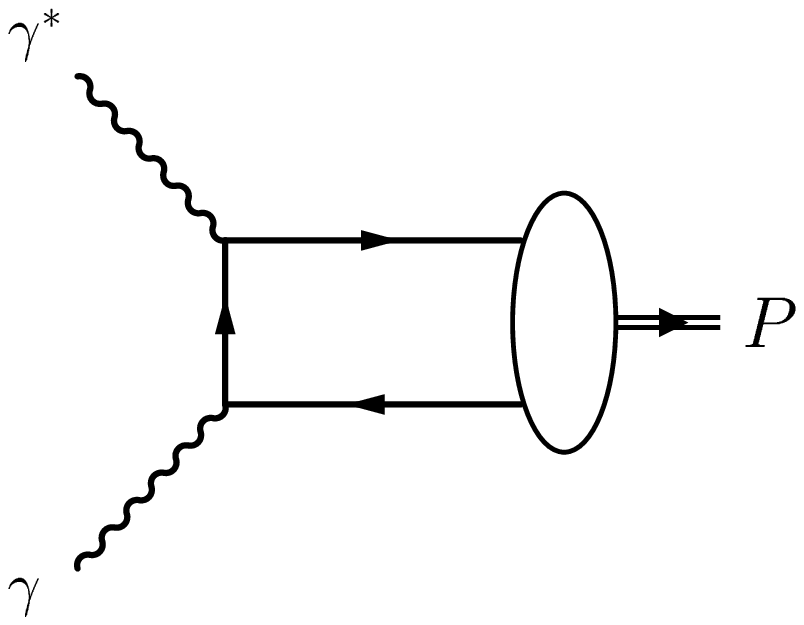}
\hspace{1em}
\includegraphics[width=4.0cm, bb=200 460 482 660,clip=true]{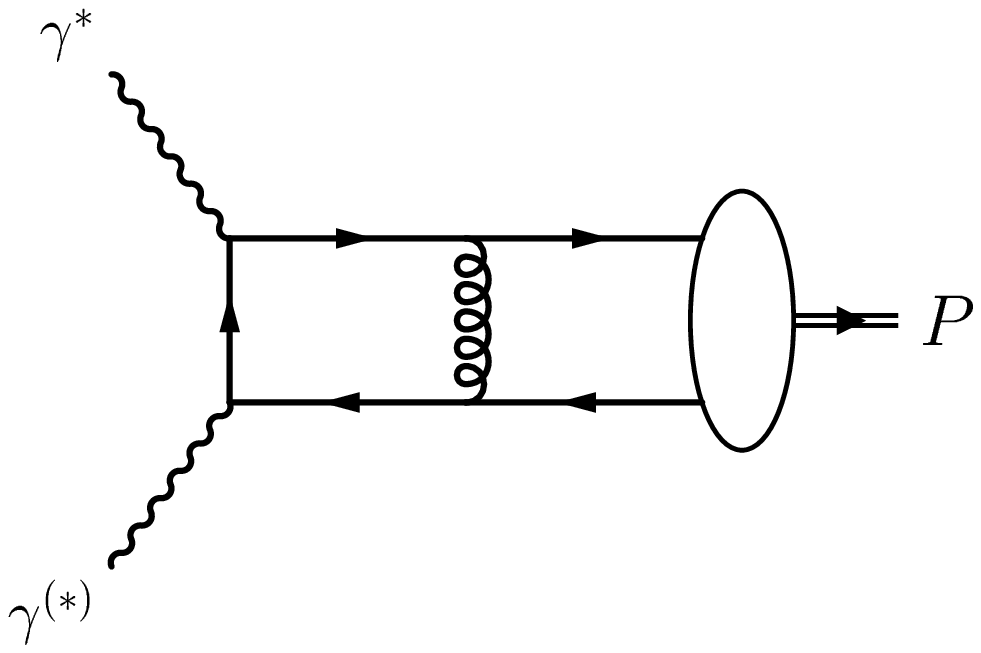}
\hspace{1em}
\includegraphics[width=4.0cm, bb=200 470 482 660,clip=true]{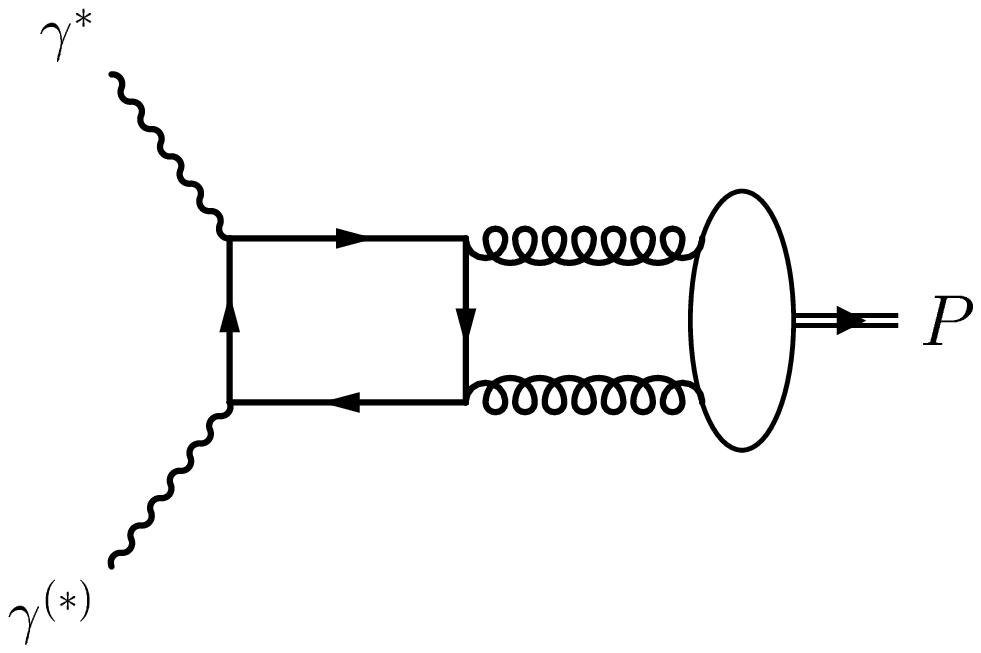}
\end{center}}
\caption{Sample Feynman graphs contributing to the transition form factors}
\label{fig:graphs}
\end{figure}

The $\gamma^*(q,\mu)\, \gamma^{(*)}(q^\prime,\nu)\to P(p)$ vertex is
parameterized as 
\ba 
\Gamma^{\mu\nu}=i e_0^2 F_{P\gamma}(\ov{Q},\omega)
\varepsilon^{\mu\nu\alpha\beta}\, q_{\alpha}\, q^\prime_{\beta}\,,
\ea
where $Q^2=-q^2 \geq 0$, $Q^{\prime 2}=-q^{\prime 2}\geq 0$ and
\be
\ov{Q}^2=\frac12 (Q^{2}+Q^{\prime 2})\,, \qquad
\omega=\frac{Q^{2}-Q^{\prime 2}}{Q^{2}+Q^{\prime 2}}\,.
\ee
Due to Bose symmetry the transition form factor is symmetric in $\omega$.
To leading-twist NLO accuracy the transition
form factor reads ($P=\eta,\eta^\prime$)
\ba
F_{P\gamma^*}&=& \frac{{2}}{3\sqrt{3}\,\ov{Q}^2}\, \int_{-1}^1\,
\frac{d\xi}{1-\xi^2\omega^2}\,\left\{
\left[\frac{f_P^{(8)}}{2\sqrt{2}}\, \Phi_{P8}(\xi,\mu_F) 
+ f_P^{(1)}\,  \Phi_{P1}(\xi,\mu_F)\right] \right.\nn\\
&\times&\left.\left[1+\frac{\als(\mu_R)}{4\pi}
      {\cal K}_q(\omega,\xi,\ov{Q}^2)\right]
   +f_P^{(1)}\,  \Phi_{Pg}(\xi,\mu_F)\,\frac{\als(\mu_R)}{4\pi}
  {\cal K}_g(\omega,\xi,\ov{Q}^2)\right\}\,. 
\label{eq:ff}
\ea
The NLO hard scattering kernels, ${\cal K}$, are calculated from the Feynman
graphs shown in Fig.\ \ref{fig:graphs}. The results - in the
$\overline{\rm MS}$ scheme - can be found in the
literature, see for instance \cite{DKV1,kornelija,braaten}. The decay
constants, $f^{(i)}_P$, are defined  by matrix elements of $SU(3)_F$
singlet and octet axial vector currents:
\be
\langle 0|J^{(i)}_{5\mu}|P(p)\rangle = i f_P^{(i)}\, p_\mu\,.
\ee   
The singlet decay constant $f_P^{(1)}$ depends on the scale
\cite{kaiser} but the
anomalous dimension controlling it is of order $\als^2$. In a NLO
calculation this effect is to be neglected for consistency.
Note that the octet part of \req{eq:ff} also holds for the
$\pi-\gamma$ form factor with the obvious replacement $\Phi_{P8} \to
\Phi_\pi$, $f_P^{(8)}\to \sqrt{3} f_\pi$.

Of particular interest is the limit $\omega\to 0$. Inserting the
Gegenbauer expansion \req{eq:gegenbauer} into \req{eq:ff}, one finds
that the Gegenbauer coefficients of the quark and gluon \da
s first appears at order $\omega^n$ \cite{DKV1}. Hence, one obtains
the prediction
\be
F_{P\gamma^*}(\ov{Q}^2,\omega)= \frac{\sqrt{2}}{3\sqrt{3}}\, 
    \frac{f_P^8 +2\sqrt{2} f_P^1}{\ov{Q}^2}\,
                  \Big[1-\frac{\als}{\pi}\Big] + {\cal O}(\omega^2,\als^2)\,.
\label{eq:omlimit}
\ee
Since the decay constants are known to amount to 
\be
f_\eta^{(8)}=1.17 f_\pi\,, \quad f_\eta^{(1)}=0.19 f_\pi\,,  \quad
f_{\eta^\prime}^{(8)}=-0.46 f_\pi\,, \quad f_{\eta^\prime}^{(1)}=1.15 f_\pi\,,
\label{eq:decay}
\ee
with a accuracy of about $5\%$ \cite{FKS1}, \req{eq:omlimit} is a
parameter-free prediction of QCD to leading-twist accuracy. Its 
theoretical status is comparable to that of the Bjorken sum rule \cite{kataev}
\be
\int_0^1 dx\, \big[g_1^p(x)-g_1^n(x)\big] = \frac16\, \frac{G_A}{G_V}
\, \Big[1 - \frac{\als}{\pi}-3.583 (\frac{\als}{\pi})^2 -20.215
  (\frac{\als}{\pi})^3 + \cdots \Big]\,,
\ee
and a few other observables among which is the famous result for the
cross section ratio of $e^+e^-$ annihilation into hadrons and into a
pair of muons. It is known \cite{MMK} that the perturbative series of
the transition form factors are identical to that of the Bjorken sum 
rule. The prediction \req{eq:omlimit} well deserves experimental
verification but there is no data as yet.

The real photon case, $\omega=1$, is another interesting limit. Here
data is available \cite{CLEO,L3} from which information about the \da s 
can be extracted. For the case of the pion such analyses have been
carried through immediately after the advent of the CLEO data in Ref.\
\cite{raulfs,rady} and, recently, in much greater detail in \cite{DKV1}. 
The $\eta$ and $\eta^\prime$ data have been analyzed within the modified
pertrubative approach in \cite{feldmann} and to leading twist NLO
accuracy in \cite{kornelija}. Since the present quality of the data
does not suffice to determine all six \da s, one has to simplify
matters and employ an $\eta-\eta^\prime$ mixing scheme in order to
reduce the number of free parameters. Since in hard processes only
small spatial quark-antiquark separations are of relevance, it is
sufficiently suggestive to embed the particle dependence and the
mixing behaviour of the valence Fock components solely into the decay
constants which play the role of wave functions at the
origin. Following \cite{FKS1,feldmann}, one may therefore take
\be
       \Phi_{Pi}=\Phi_i\,, \qquad  \Phi_{Pg}=\Phi_g\,.
\label{eq:indep}
\ee
%The relative strength of the gluonic $\eta$ and $\eta^\prime$ \da s,
%is thus controlled by the singlet decay constants {\bf spaeter?}. 
This assumption is further supported by the observations made in
\cite{feldmann} that, as is the case for the pion
\cite{DKV1,raulfs,rady} the quark \da s are close to the asymptotic
form, $\Phi_{AS}$, for which the particle independence \req{eq:indep} 
holds trivially. The analysis is further simplified by truncating the
Gegenbauer series in \req{eq:gegenbauer} at $n=2$. The coefficients
$B_2^{(i)}$, acting for all others, parameterize the deviations from
the asymptotic form of the \da s. Clearly, this is a serious assumption  
(note that to LO accuracy the transition form factors only fix the sum
$1+\sum B_{n}^{(i)}$) but in view of the large experimental errors as
well of the limited range of momentum transfer in which data is
available, one is forced to do so. Truncation at $n=4$ does not lead
to reliable results, all contributing Gegenbauer coefficients are
highly correlated. A fit to the CLEO and L3 data provides
\be
B_2^{(8)}(\mu_0)=-0.04\pm 0.04\,, \quad B_2^{(1)}(\mu_0)=-0.08\pm 0.04\,, 
\quad B_2^{(g)}(\mu_0)= 9\pm 12\,,
\label{eq:parameters}
\ee
where the following scales have been chosen: $\mu_0=1\gev$, $\mu_F=Q$,
$\mu_R=Q/\sqrt{2}$. The use of $\mu_F=Q/\sqrt{2}$ instead leads to
values of the Gegenbauer coefficients which agree with those quoted in
\req{eq:parameters} almost perfectly. For comparison, $B^\pi_{2}$
takes a value of $-0.06\pm 0.03$ as determined in \cite{DKV1}. 

\begin{figure}
\parbox{\textwidth}{
\begin{center}
\includegraphics[width=5.9cm]{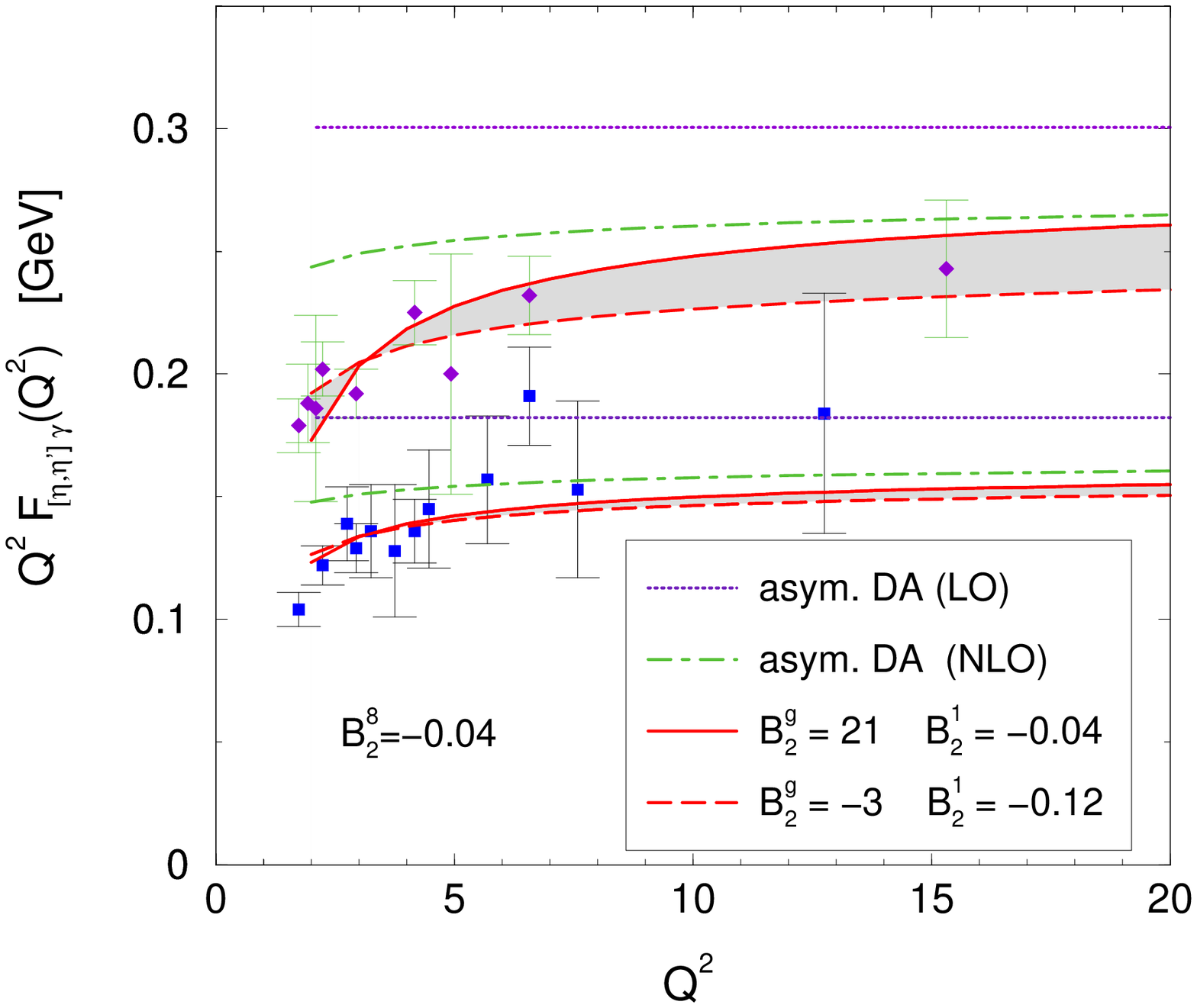}
%\hspace{1em}
%\includegraphics[width=5.5cm]{corr.eps}
\hspace{1em}
\includegraphics[width=5.9cm]{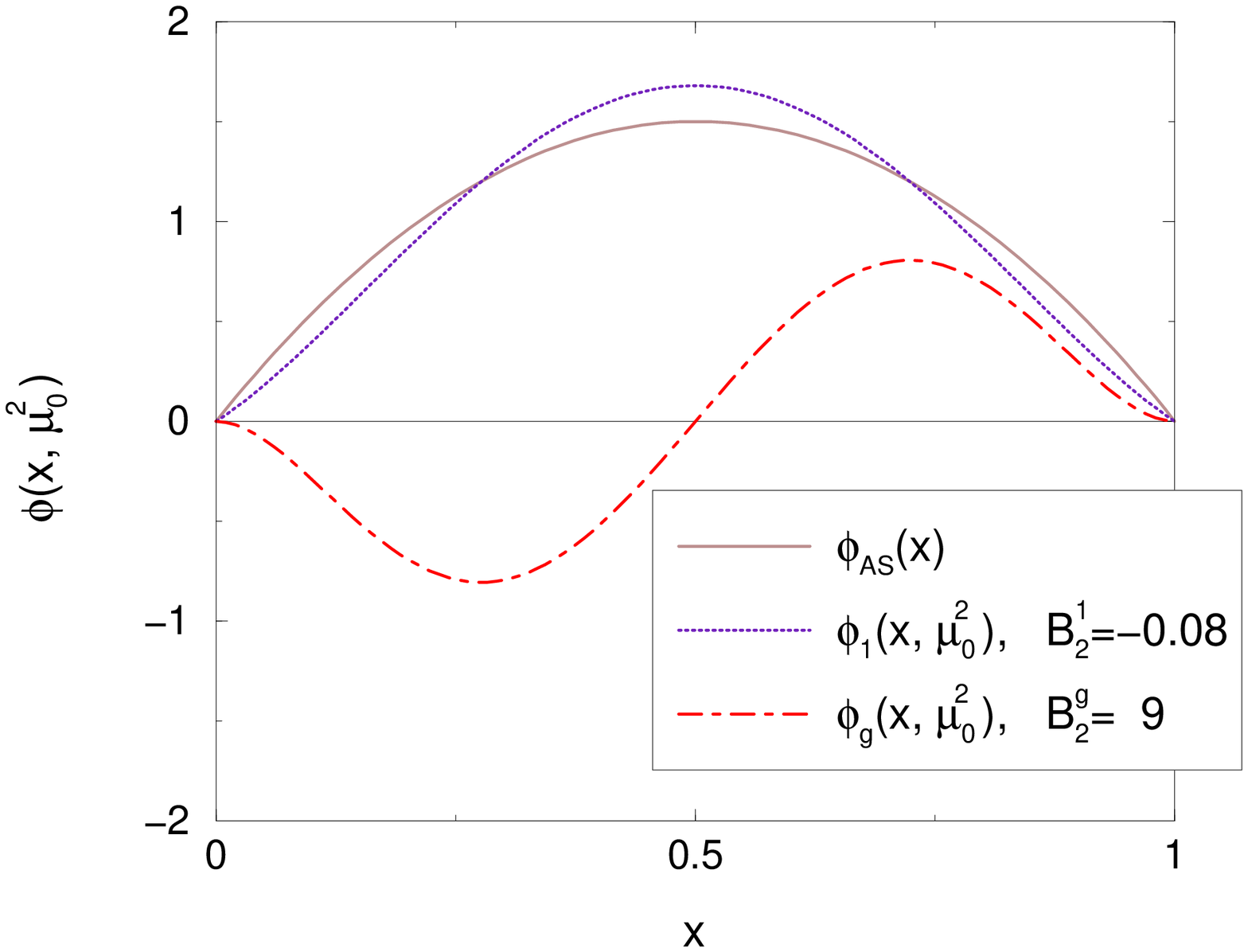}
\end{center}}
\caption{Left:The scaled $P\gamma$ transition form factor vs.\ $Q^2$. Data
taken from \protect\cite{CLEO,L3}; rombs represent the $\eta^\prime$ data,
  squares the $\eta$ ones. Right: The flavour-singlet and gluon \da s
  at the scale $\mu_0=1 \gev^2$}
\label{fig:results}
\end{figure}

The fit is compared to the data in Fig.\ \ref{fig:results}. The
insensitivity of the $\eta-\gamma$ transition form factor to the
gluonic \da{} is clearly seen which comes about as a consequence of
the smallness of $f_\eta^{(1)}$, see \req{eq:decay}. Although
the present data are compatible with a leading-twist analysis as Fig.\
\ref{fig:results}, the existence of power and/or higher-twist
corrections cannot be excluded. This is a source of theoretical
uncertainties in the results \req{eq:parameters}. Thus, for instance,
the use of the modified perturbative approach in which quark
transverse degrees of freedom and Sudakov suppressions are taken into
account, leads to good agreement with experiment for the asymptotic
\da s \cite{feldmann}.

Within errors the quark Gegenbauer coefficients for the octet and
singlet case agree with each other and with the pion one. This implies
not only approximate flavour symmetry but also the approximate
validity of the OZI rule which is a prerequisite of the quark-flavour 
mixing scheme advocated for in \cite{FKS1}. Although the face value of 
$B_2^{(g)}$ is huge as compared to that of $B_2^{(1)}$ the gluonic 
\da{} itself is not large as can be seen from Fig.\ \ref{fig:results}, its
$x\leftrightarrow 1-x$ asymmetry and the numerical factors in
\req{eq:gegenbauer} keep it small. Moreover since it only contributes
to NLO its impact on the transition form factors is small resulting in
large errors. In order to obtain more precise information on the
gluonic \da{} additional constraints from other reactions are required. 
The inclusive decay $\Upsilon (^1S)\to \eta^\prime X$,
discussed in \cite{ali}, is one such possibility. Others are e.g.\
$B\to \pi\eta^\prime$ or $\chi_{cJ} \to \eta^\prime\eta^\prime$.
Finally it is to be emphasized that the approach presented in this
article applies to all flavour-neutral pseudoscalar mesons, e.g.\ for
the $\eta(1400)$. The properties of the valence \da s
\req{eq:gegenbauer} make it unlikely that a pseudoscalar meson
possesses pure glueball properties. A substantial $\qbq$ Fock
component is always there. For flavour-neutral scalar mesons, on the 
other hand, the situation is different. The properties of the quark
and gluon \da s are reversed \cite{chase}. A strong $gg$ Fock
component is therefore not necessarily accompanied by
strong $\qbq$ one.

\bibliographystyle{aipprocl} % if natbib is missing

%%%%%%%%%%%%%%%%%%%%%%%%%%%%%%%%%%%%%%%%%%%%%%%%%%%%%%%%%%%%%%%%%%%%%%%%%%%

\end{document}